\documentclass[aps,pre,preprint,groupedaddress]{revtex4-1}

\usepackage{graphicx,amsmath,amsfonts}
\usepackage{epstopdf}

\newcommand{\E}{\mathbb{E}}

\begin{document}


\title{Finite-size effects in a stochastic Kuramoto model}

\author{Georg A. Gottwald}
\email{georg.gottwald@sydney.edu.au}
\affiliation{School of Mathematics and Statistics, University of Sydney, NSW Sydney 2006, Australia.}

\date{\today}

\begin{abstract}
We present a collective coordinate approach to study the collective behaviour of a finite ensemble of $N$ stochastic Kuramoto oscillators using two degrees of freedom; one describing the shape dynamics of the oscillators and one describing their mean phase. Contrary to the thermodynamic limit $N\to\infty$ in which the mean phase of the cluster of globally synchronized oscillators is constant in time, the mean phase of a finite-size cluster experiences Brownian diffusion with a variance proportional to $1/N$. This finite-size effect is quantitatively well captured by our collective coordinate approach.  
\end{abstract}

\pacs{}

\maketitle

\section{Introduction}
From flavour evolution of massive neutrinos \cite{Pantaleone98}, the rhythmic activity in the brain \cite{SheebaEtAl08,BhowmikShanahan12}, to simultaneous clapping in concert halls \cite{NedaEtAl00} and power grids \cite{FilatrellaEtAl08}, networks of coupled oscillators are ubiquitous in nature and the human made world. All these examples share the same tendency of their respective oscillators to organise themselves in collective synchronised behaviour. This can be detrimental to the functioning of the network, such as in epilepsy in the brain, or may be desired, as in the smooth running of power supply in power grids. The {\em{drosophila melanogaster}} of the study of coupled oscillators has been for the last 40 years the celebrated first-order Kuramoto model \cite{Kuramoto,OsipovEtAl,PikovskyEtAl,Strogatz00,AcebronEtAl05,ArenasEtAl08,DoerflerBullo14,RodriguesEtAl16}, despite only representing an idealized case of global all-to-all sine-coupling between oscillators. In neuroscience it has been used to model oscillatory activity in the cortex \cite{BreakspearEtAl10}, to provide a mechanism for low-frequency fluctuations of the brain's resting-state as seen in fMRI data \cite{CabralEtAl11} and to understand the generation of $\delta$- and $\theta$-waves during anaesthesia \cite{SheebaEtAl08}. Naturally occurring systems are subject to random external fluctuations. In particular, it was realized that the presence of noise in nervous systems has a profound impact of the brain's structure and function \cite{FaisalEtAl08, GoldwynSheaBrown11}. To study the effect of external noise on phase oscillators and on their collective behaviour, stochastic Kuramoto models were introduced, see for example \cite{Sakaguchi88, SonnenscheinSchimanskyGeier12, SonnenscheinSchimanskyGeier13}. In a recent series of mathematical work, the effect of external noise in a Kuramoto model was studied, where it was found that in finite networks the mean phase of the synchronized cluster of oscillators exhibits Brownian motion with variance proportional to $1/N$ \cite{BertiniEtAl10,LuconPoquet15,GiacominPoquet15,Lucon15,LuconStannat16}. There are in fact two types of finite size effects in the stochastic Kuramoto model. First, there is a sampling error of the native frequencies, which results in a non-zero mean frequency of the order $1/\sqrt{N}$. This induces a drift which is also present in the deterministic Kuramoto model. The second finite-size effect are noise-induced fluctuations of the global mean phase with variance proportional to $1/N$; oscillators form a synchronized cluster which is then as a whole subjected to diffusive behaviour along the neutral direction associated with the invariance of the Kuramoto model with respect to translations of the mean phase.\\
Capturing the behaviour of finite-size networks is one of the most challenging puzzles in the study of the collective behaviour of coupled phase oscillators. Current theory mostly relies on the thermodynamic limit of infinite networks \cite{DorogovtsevEtAl08,OttAntonson08,RosenblumPikovsky15} and cannot reproduce the observed behaviour of finite-size networks. One way to approach finite-size effects is to obtain information about multi-oscillator correlations in a kinetic theory approach \cite{HildebrandtEtAl07,BuiceChow07,HongEtAl07,Tang11}. We employ here the much simpler and easier to calculate collective coordinate  approach proposed in \cite{Gottwald15}, leading to a low-dimensional description of the stochastic Kuramoto model.\\

The collective behaviour of coupled oscillators such as synchronisation behaviour suggests that the dynamics of complex systems may (at least in certain cases) be described by a low dimensional dynamical system. 
Recently we introduced a new framework for such a model reduction based on collective coordinates which can deal with the physically relevant case of finite network sizes \cite{Gottwald15}. This collective coordinate approach has since been successfully applied to derive optimal synchronization design strategies and optimal synchrony network topologies \cite{PintoSaa15,BredeKalloniatis16}. Here we will use collective coordinates to accurately describe the finite-size effects of a stochastic Kuramoto model, in particular the diffusion of the mean phase.\\

The paper is organised as follows. In Section~\ref{s.SK} we introduce the stochastic Kuramoto model. We then perform the collective coordinate approach in Section~\ref{s.cc} and present numerical results in Section~\ref{s.numerics} illustrating the ability of our approach to capture the finite size effects of this model. We conclude in Section~\ref{s.discussion} with a discussion.

\vspace{-0.6cm}

\section{Stochastic Kuramoto model}
\label{s.SK} 
We consider the stochastic Kuramoto model for $N$ phase oscillators $\varphi_i$ 
\begin{align}
d \varphi_i = \omega_i\, dt + \frac{K}{N}\sum_{i=1}^N\sin{(\varphi_j-\varphi_i)}\, dt + \sigma dB_{i,t}\, ,
\label{e.SK}
\end{align}
where $\omega_i$ are their native frequencies and $\sigma^2$ is the homogenous variance of the noise $dB_t\sim{\mathcal{N}}(0,t)$ \cite{Sakaguchi88,LuconPoquet15}. We consider here the case of an all-to-all coupling topology where each oscillator is coupled to all the other oscillators.  Without loss of generality we assume here that the native frequencies are drawn from a distribution with mean zero. For a finite ensemble of $N$ native frequencies, however, the mean will not be exactly zero but will have sample errors of size $1/\sqrt{N}$.

The level of synchronisation is often characterised by the order parameter \cite{Kuramoto}
\begin{align*}
r(t)=\frac{1}{N}|\sum_{j=1}^Ne^{i\varphi_j(t)}|\; ,
\end{align*} 
with $0\le r \le 1$. In practice, one determines its long-time average ${\bar{r}}$.
In the case of full synchronisation with $\varphi_i(t)=\varphi_j(t)$ for all pairs $i,j$ and for all times $t$ we obtain $\bar{r}=r=1$. In the case where all oscillators behave independently with random initial conditions $\bar{r} = \mathcal{O}(1/\sqrt{N})$ indicates incoherent phase dynamics; values in between indicate partial coherence.


\section{Collective Coordinates}
\label{s.cc}
\begin{figure}[htbp]
\centering
\includegraphics[width = 0.5\columnwidth, height = 6cm]{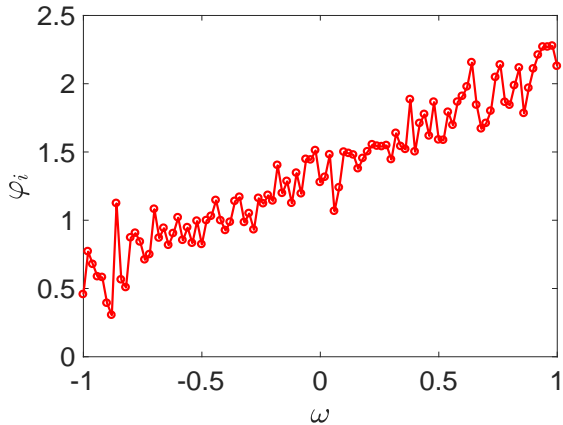} 
\caption{Snapshot of the phases $\varphi_i$ as a function of the sorted native frequencies $\omega_i$ for the stochastic Kuramoto model (\ref{e.SK}) with native frequencies drawn from the uniform distribution ${\mathcal{U}}\sim[-1,1]$. Here $N=100$ and $\sigma=0.2$ at $K=1.49$. The average linear functional form $\varphi_i=\alpha \, \omega_i + \psi$ is clearly seen.}
\label{f.phis}
\end{figure}

Collective coordinates have been introduced in the context of conservative nonlinear wave dynamics \cite{Scott}. The theory has then been extended to dissipative systems such as reaction diffusion systems \cite{GottwaldKramer04,MenonGottwald05,MenonGottwald07,MenonGottwald09,CoxGottwald06} and recently to phase oscillators \cite{Gottwald15}. Here we extend the method to the stochastic Kuramoto model.\\ In order to find a dimension reduced description of the stochastic Kuramoto model (\ref{e.SK}) we make the following ansatz
\begin{align}
\varphi_i(t)=\alpha(t)\, \omega_i + \psi(t)\, ,
\label{e.cc}
\end{align}
which defines the collective coordinates $\alpha(t)$ and $\psi(t)$. The collective coordinates will be stochastic processes evolving to some as yet unspecified dynamics
\begin{align}
d\alpha & = a_\alpha(\alpha,\psi)\, dt + \sigma_{\alpha\alpha}\, dW_1 + \sigma_{\alpha \psi}\, dW_2\, ,
\label{e.alpha}
\\
d\psi &=  a_\psi(\alpha,\psi)\, dt + \sigma_{\psi\alpha}\, dW_1 + \sigma_{\psi\psi}\, dW_2\, ,
\label{e.f}
\end{align}
where the diffusion coefficients may also be functions of the collective coordinates $\alpha$ and/or $\psi$ and $W_{1,2}$ are one-dimensional Brownian motions. Solutions with $d\E[\alpha]=0$, where the expectation value is taken over the invariant measure, correspond to phase-locked solutions with mean-phase $\psi(t)$. The existence of such solutions corresponds to a synchronised state. Before deriving analytical expressions for the drift and diffusion coefficients in (\ref{e.alpha})-(\ref{e.f}), let us first motivate the collective coordinate ansatz (\ref{e.cc}). For $\psi= 0$, the ansatz has been used in \cite{Gottwald15} to describe the behaviour of the deterministic Kuramoto model with $\sigma =0$. Figure~\ref{f.phis} shows a snapshot of the phases $\varphi_i$ as a function of the ordered native frequencies for the stochastic Kuramoto model (\ref{e.SK}). The phases of the entrained oscillators still obey the ansatz (\ref{e.cc}) on average. The ansatz can be motivated by considering large coupling strength $K\gg 1$. The averaged Kuramoto model (\ref{e.SK}) with can be cast as $\omega_i = - K r \sin(\psi-\varphi_i)$ introducing the mean phase $\psi$ \cite{Kuramoto}. Expanding $\varphi_i=\psi +\arcsin(\omega_i/(rK))$ in $1/K$ for large coupling strength yields up to first order $\varphi_i = \psi + \omega_i/(rK)$. Since the Kuramoto model is invariant under constant phase shifts we may set $\psi=0$ leading to ansatz (\ref{e.cc}) with $\psi\equiv 0$. To capture, however,  the finite size effects of the stochasticity of the Kuramoto model with $\sigma \neq 0$ we introduce a time dependent mean phase $\psi(t)$, allowing for diffusive Brownian behaviour along the group orbit associated with this continuous symmetry.\\ 

To determine the drift and diffusion of the evolution equations (\ref{e.alpha})--(\ref{e.f}) for the collective coordinates we require that the error made by the collective coordinates (\ref{e.cc}) is minimized. We introduce the error, made by restricting the solution to the subspace defined by the ansatz (\ref{e.cc}), 
\begin{align*}
d\mathcal{E}_{\rm{coll}} 
= & 
( a_\alpha \omega_i + a_\psi - \omega_i - \frac{K}{N} \sum_{j=1}^N \sin(\alpha(\omega_j-\omega_i)))\, dt \\
&+ 
\left(\sigma_{\alpha\alpha}\omega_i+\sigma_{f\alpha}\right) dW_1 + \left(\sigma_{\alpha f}\omega_i + \sigma_{ff}\right) dW_2 \\
&-\sigma dB_i\, .
\end{align*}
Note that there are no It\^o-corrections since $\varphi_i$ is linear in the collective coordinates $\alpha$ and $\psi$. Minimiziation of the error $\mathcal{E}_{\rm{coll}}$ is achieved by assuring that it is orthogonal to the restricted subspace spanned by (\ref{e.cc}). We therefore require that $\mathcal{E}_{\rm{coll}}$ be orthogonal to the tangent space of the solution manifold (\ref{e.cc}) which is spanned by $\partial \varphi_i/\partial \alpha = \omega_i$ and $\partial \varphi_i/\partial  \psi=1$. Projecting the error then yields the desired drift and diffusion coefficients. We obtain for the drift coefficients of the collective coordinate evolution equation 
\begin{align}
a_\alpha &= 1 + \frac{1}{H^2}\frac{K}{N^2} \sum_{i=1}^N\omega_i \sum_{j=1}^N \sin(\alpha(\omega_j-\omega_i))\, ,
\label{e.cc_drift_alpha}
\\
a_\psi &= -(1-a_\alpha)\bar\Omega\; ,
\label{e.cc_drift_psi}
\end{align}
where we introduced the sample mean and variances
\begin{align*}
\bar \Omega = \frac{1}{N}\sum_{j=1}^N\omega_j\,,  \quad  S^2 = \frac{1}{N}\sum_{j=1}^N\omega_j^2\, ,
\end{align*}
and defined $H^2 = S^2-\bar\Omega^2$. The drift coefficient $a_\alpha$ (\ref{e.cc_drift_alpha}) is exactly the same as for the deterministic Kuramoto model \cite{Gottwald15}. The projections of the stochastic terms are written as
\begin{align*}
\sigma_{\rm{coll}}\,d{\mathcal{W}} = \sigma_{\rm{K}}\,d{\mathcal{B}}
\end{align*}
with the Brownian motions ${\mathcal{W}}=(W_1,W_2)^T$ and ${\mathcal{B}}=(B_1,B_2,\cdots,B_N)^T$ and covariance matrices
\begin{align*}
\sigma_{\rm{coll}}=
\left(
\begin{matrix}
\sigma_{\alpha \alpha}S^2+\sigma_{\psi\alpha}\bar\Omega & \sigma_{\alpha \psi}S^2+\sigma_{\psi\psi}\bar\Omega\\
\sigma_{\alpha \alpha}\bar\Omega + \sigma_{\psi\alpha} & \sigma_{\alpha \psi}\bar\Omega+\sigma_{\psi\psi}
\end{matrix}
\right)
\end{align*}
and
\begin{align*}
\sigma_{\rm{K}}=
\frac{\sigma}{N}
\left(
\begin{matrix}
\omega_1&\omega_2 & \cdots & \omega_N\\
1&1 & \cdots & 1\\
\end{matrix}
\right)\, .
\end{align*}
To assure that the collective coordinates capture the statistics of the stochastic Kuramoto model we require that the diffusivities of the associated Fokker-Planck equations coincide, i.e. we require 
$\frac{1}{2} \sigma_{\rm{K}} \sigma_{\rm{K}}^T 
= 
\frac{1}{2}  \sigma_{\rm{coll}} \sigma_{\rm{coll}}^T $.
The matrix $\sigma_{\rm{coll}}$ is then found as a square root matrix 
leading finally to the diffusion coefficients of the collective coordinate evolution equation 
\begin{align*}
\sigma_{\alpha\alpha} &= \sigma_{\rm{cc}} \,(1+\frac{1}{H})\, , \;\;\sigma_{\alpha \psi} =-\sigma_{\rm{cc}} \, \frac{\bar\Omega}{H}=\sigma_{\psi \alpha} \\
\sigma_{\psi \psi} &= \sigma_{\rm{cc}} \,(1+\frac{S^2}{H})\, ,
\end{align*}
where
\begin{align*}
\sigma_{\rm{cc}} = \frac{\sigma}{\sqrt{N}\sqrt{S^2+1+ 2 H}}\, .
\end{align*}
This concludes the derivation of the dynamics of the collective coordinates. We summarize
\begin{align}
d\alpha& = \big( 
1 + \frac{1}{H^2}\frac{K}{N^2} \sum_{i=1}^N\omega_i \sum_{j=1}^N \sin(\alpha(\omega_j-\omega_i))
\big)\, dt
\nonumber
\\
&\quad
+ \frac{\sigma}{\sqrt{N}\sqrt{S^2+1+ 2 H}}\,( (1+\frac{1}{H})\, dW_1 -  \frac{\bar\Omega}{H}\,dW_2)
\label{e.dalpha}
\\ 
d\psi &= -(1-a_\alpha)\bar\Omega\, dt
\nonumber
\\
& \quad 
- \frac{\sigma}{\sqrt{N}\sqrt{S^2+1+ 2 H}}\,(\frac{\bar\Omega}{H}\,dW_1 + (1+\frac{S^2}{H})\, dW_2)\, .
\label{e.dpsi}
\end{align}
These reduced equations for the collective coordinates allow us to study the onset of synchronisation by analysing a reduced two-dimensional problem and furthermore allow for the analysis of finite network size $N$. Note that the noise is additive and the systems of stochastic differential equations (\ref{e.dalpha})-(\ref{e.dpsi}) is a skew-product system and hence the joint probability factorizes as $\rho(\alpha,\psi)=\rho_\psi(\psi|\alpha)\rho_\alpha(\alpha)$. The dynamics of $\alpha$ allows for a stationary distribution $\rho_\alpha(\alpha)$ around the deterministic mean $\alpha^\star$ which is the solution to $a_\alpha(\alpha^\star) = 0$. Solutions with $d\E[\alpha]=0$, i.e. $a_\alpha = 0$, where the expectation value is taken over the invariant measure, correspond to phase-locked synchronized state with mean-phase $\psi(t)$. The collective coordinate equations (\ref{e.dalpha})-(\ref{e.dpsi}) show that in the thermodynamic limit the Brownian driving force disappears and the synchronization behaviour is given by the deterministic Kuramoto model with $\rho_\alpha(\alpha)=\delta(\alpha-\alpha^\star)$. For finite $N$, the synchronization is delayed for non-zero noise variance $\sigma^2$; this can be seen in the thermodynamic limit, where the order parameter $r\exp(i\psi)=\sum_{i=1}^N \exp(i\varphi_i)/N$ can be readily evaluated by averaging over the frequency distribution function. For uniformly distributed native frequencies, for example, we obtain, using the collective coordinate ansatz (\ref{e.cc}), $r = \sin(\alpha)/\alpha$. 
In the synchronized state, averaging over $\rho_\alpha(\alpha)$ yields $r =r_{\rm{det}}= \sin(\alpha^\star)/\alpha^\star$ in the deterministic case, and $r =r_{\rm{stoch}}= \int \sin(\alpha)/\alpha\,  \rho_\alpha(\alpha)d\alpha$ in the stochastic case. For synchronized states with sufficiently small $\alpha^\star$ the negative curvature of $r(\alpha)$ renders $r_{\rm{stoch}}<r_{\rm{det}}$, implying that noise delays the onset of synchronization to higher coupling strengths $K$. This is confirmed in numerical simulations and holds as well for finite networks.\\ 

Concentrating on the diffusive behaviour of the mean phase--our main aim here--we average the evolution equation (\ref{e.dpsi}) for the mean phase $\psi$ over the invariant density $\rho_\alpha$, and obtain
\begin{align}
\psi(t) = -\bar\Omega dt + \frac{\sigma}{\sqrt{N}}\frac{S}{\sqrt{S^2-{\bar\Omega}^2}}\, W_t\, .
\label{e.psi_sync}
\end{align}
Subtracting the mean drift, caused by the sampling error $\bar\Omega \neq 0$, the mean-square displacement of $\psi$ is found to be
\begin{align}
\Delta_{\rm{MSQ}} = \frac{\sigma^2}{N}\frac{S^2}{S^2-\bar\Omega^2}\, t\, .
\label{e.MSQ}
\end{align}
This shows that if the randomly chosen native frequencies have small mean frequency $\bar\Omega\ll 1$, the collective motion of synchronized oscillators exhibits finite-size induced Brownian motion with a mean-square displacement inversely proportional to $N$.

\section{Numerical results}
\label{s.numerics}
We now present numerical results illustrating finite-size effects of the dynamics of the stochastic Kuramoto model, and show how the collective coordinate approach is able to quantitatively capture those effects. We consider here a network with $N=100$ oscillators with native frequencies drawn from the uniform distribution ${\mathcal{U}}\sim[-1,1]$. The critical coupling strength $K_c$ above which the oscillators experience global synchronization and organize themselves into a single partially synchronized cluster is found in the noiseless case $\sigma=0$ as $K_c=1.285$ (estimated by the collective coordinate approach accurately up to $2\%$). We show here results for $K=1.39$  which corresponds to a synchronized state with order parameter $\bar r=0.84$ for $\sigma=0.2$.\\ In particular, we focus here on the dynamics of the centre of the single synchronized cluster $\psi$. We numerically estimate $\psi$ as the cluster centre 
$\psi=\varphi_{i_0}(t)$, 
where the index $i_0$ corresponds to the node with the smallest absolute values of its native frequency. We study its temporal evolution on the real line, rather than on the interval $[0,2\pi]$ to capture the unbounded drift and Brownian motion.\\  

\begin{figure}[htbp]
\centering
\includegraphics[width = 0.5\columnwidth, height = 6cm]{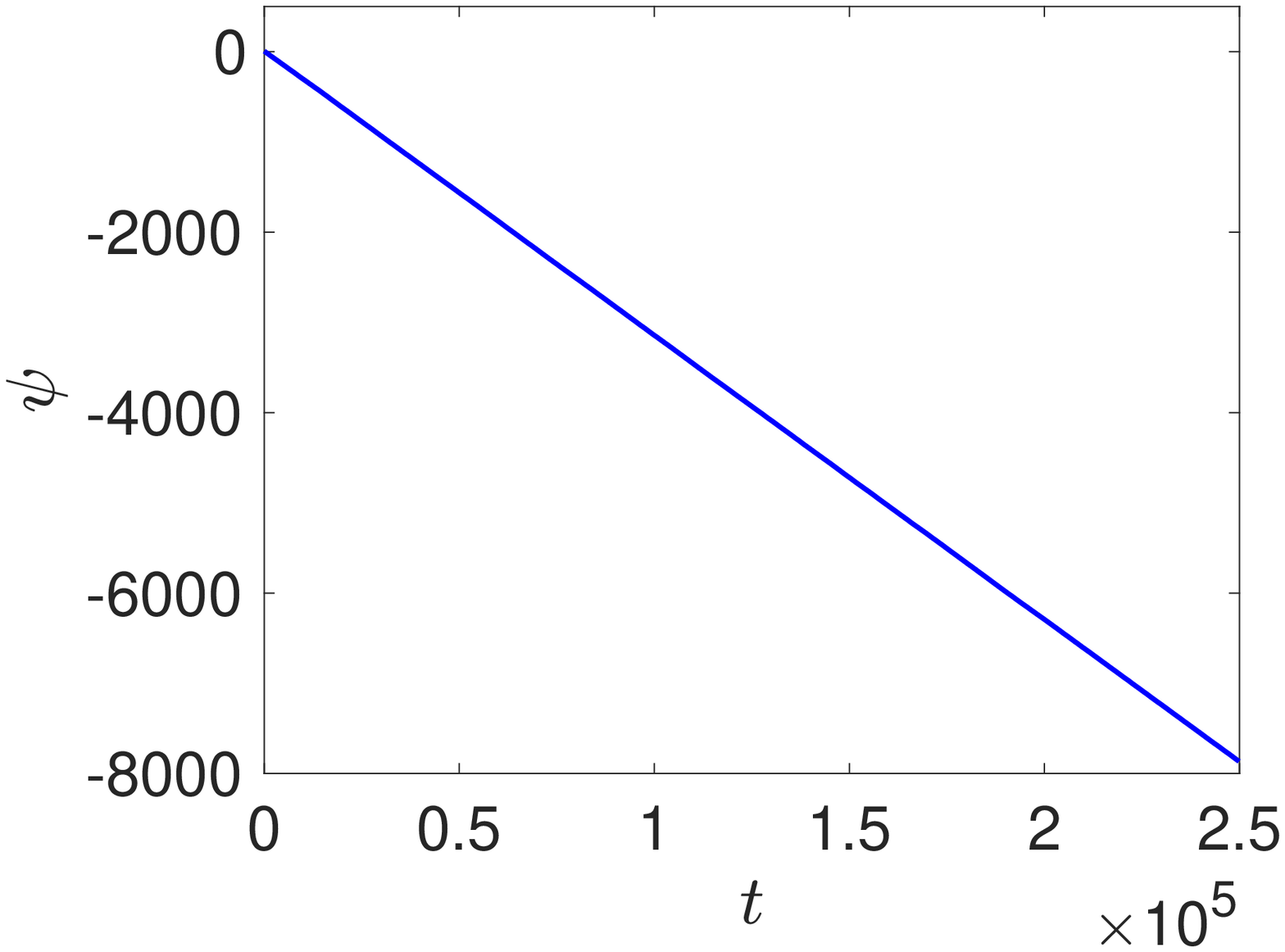}\\ 
\includegraphics[width = 0.5\columnwidth, height = 6cm]{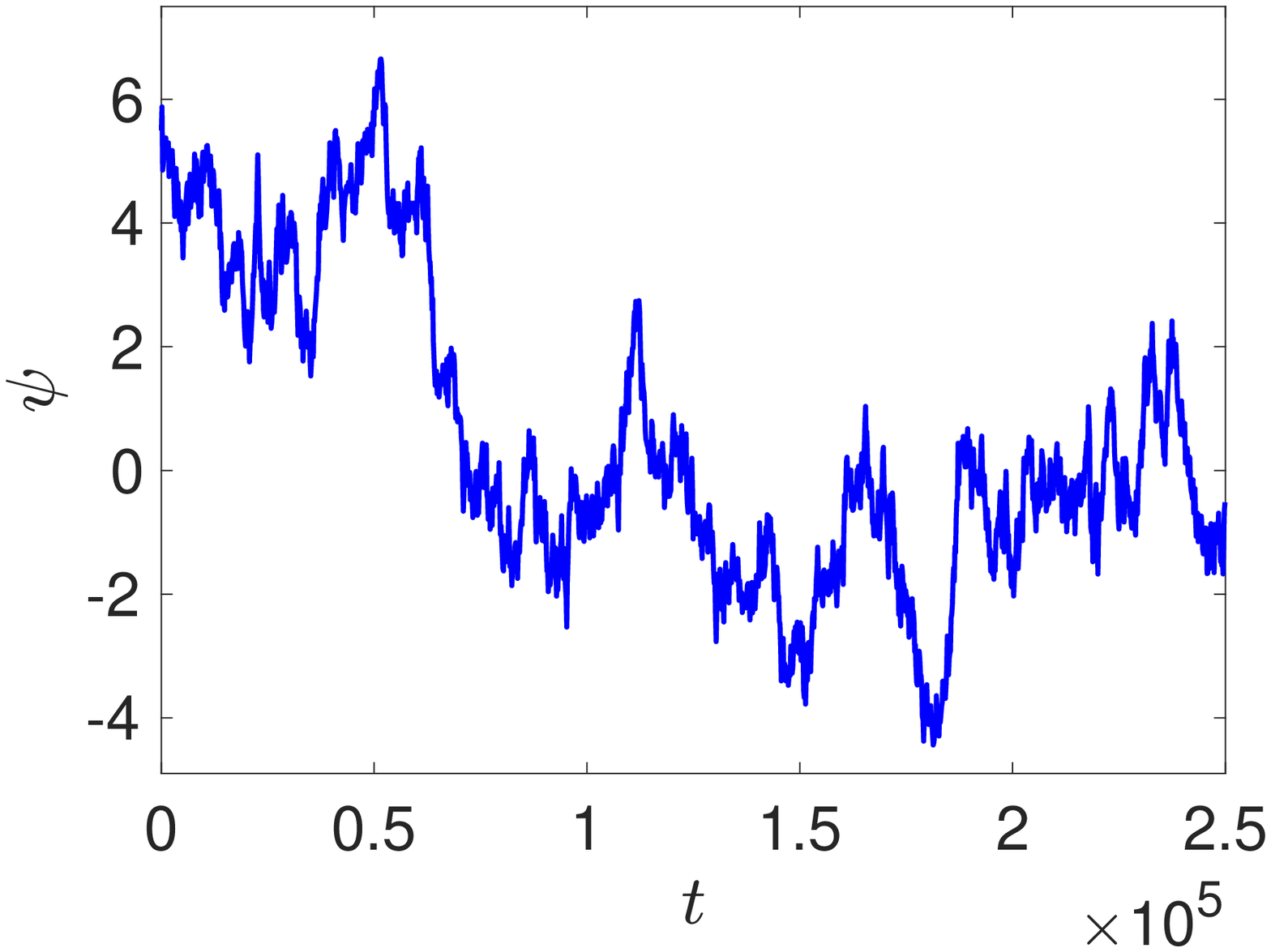} 
\caption{Center of the empirical measure of phases for a network with $N=100$ oscillators of the Kuramoto model (\ref{e.SK}) with $\sigma=0.2$ as a function of time. Simulations are shown at coupling strength $K=1.39$ for a realization of native frequencies with (a): sample mean $\bar\Omega =-0.314$ and sample variance $S^2=0.348$. (b): sample mean $\bar\Omega =10^{-17}$ and sample variance $S^2=0.343$.  }
\label{f.pi}
\end{figure}

\begin{figure}[htbp]
\centering
\includegraphics[width = 0.5\columnwidth, height = 6cm]{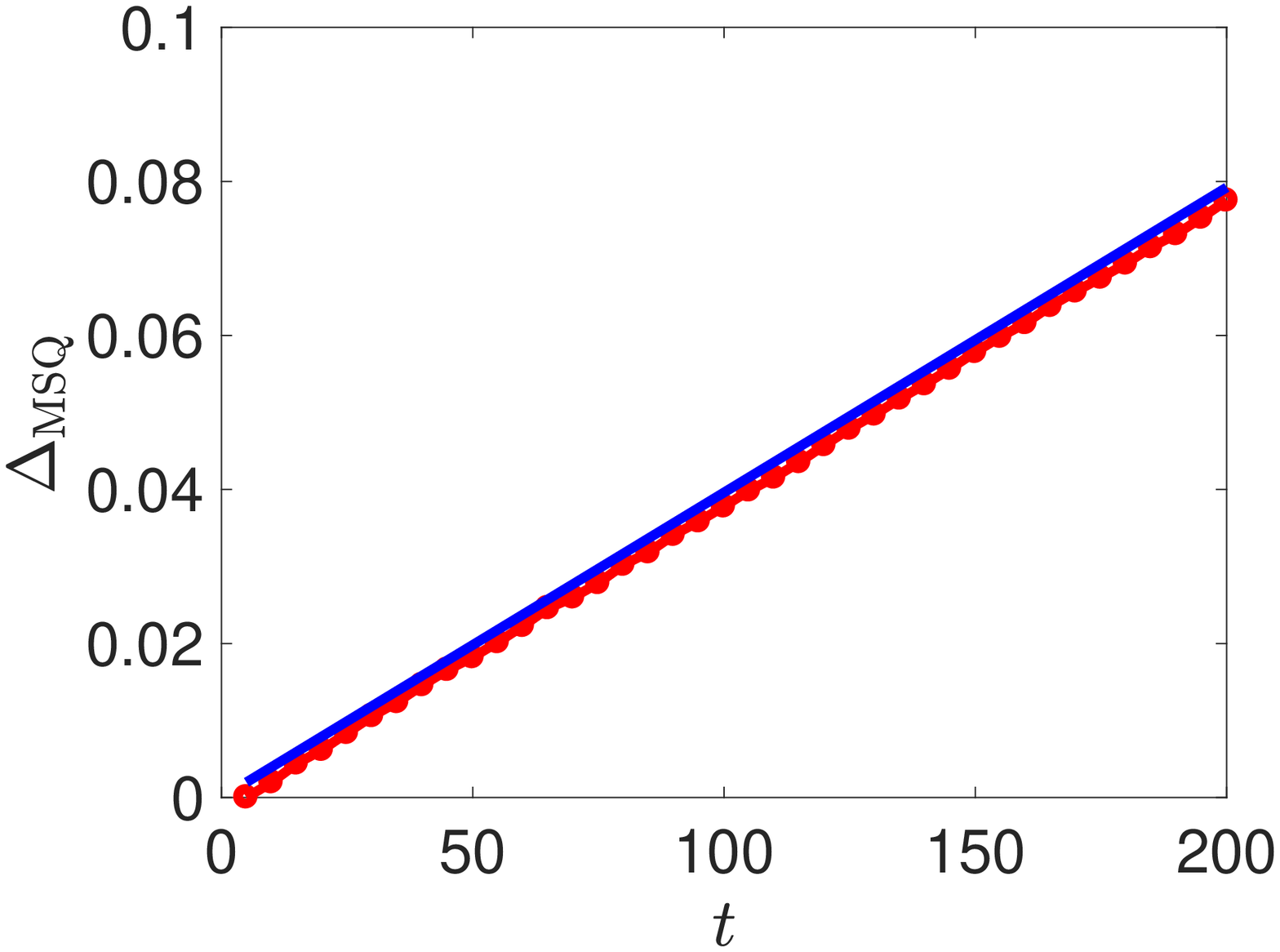} 
\caption{Mean-square displacement $\Delta_{\rm{MSQ}}$ as a function of time for a network with $N=100$ oscillators at $K=1.39$ with noise strength $\sigma=0.2$ and native frequencies with sample mean $\bar\Omega=10^{-17}$ and sample variance $S^2=0.3434$. Shown are results from a simulation of the Kuramoto model (\ref{e.SK}) (open circles (online red)) and of the analytical expression (\ref{e.MSQ}) obtained from the collective coordinate approach (continuous line (online blue)). }
\label{f.MSQ}
\end{figure}

We present results for random native frequencies with non-zero $\bar \Omega$ and with almost vanishing $\bar \Omega$. For non-vanishing $\bar \Omega$ we find a dominant linear drift of the mean phase (cf. (\ref{e.psi_sync})). This is clearly seen in Figure~\ref{f.pi}(a) where we show results from a numerical simulation of the stochastic Kuramoto model (\ref{e.SK}) for a network with native frequencies with sample mean $\bar \Omega= -0.314$ and sample variance $S^2=0.348$. The drift of the center is as expected linear in time and linear regression yields with $\psi = -0.314\, t+\psi_0=-\bar\Omega\, t + \psi_0$. More interesting is the case when the sample mean of the frequencies $\bar\Omega$ is close to zero. Figure~\ref{f.pi}(b) shows the mean phase $\psi$ of a numerical simulation of the stochastic Kuramoto model (\ref{e.SK}) for a network of oscillators with sample mean $\bar\Omega =10^{-17}$ and sample variance $S^2=0.343$. Now the mean phase clearly exhibits non-trivial stochastic behaviour. We recall that the diffusive behaviour of $\psi$ diminishes for increasing network size $N$. To quantify the stochastic motion we compute the mean-square displacement of $\psi$. This is shown in Figure~\ref{f.MSQ} where we compare the result of the numerical simulation with the analytical prediction of the collective coordinate approach (\ref{e.psi_sync}). The correspondence between the observed diffusive behaviour and the collective coordinates is remarkable with an error in the slopes, estimated by linear regression, of only $0.9\%$.

%


\section{Discussion}
\label{s.discussion}
We presented a framework for model reduction of stochastic Kuramoto systems. The approach involves the introduction of collective coordinates, parametrizing the collective behaviour of coupled oscillators on the synchronization manifold. We applied the framework here to the stochastic Kuramoto model with an all-to-all coupling topology and homogeneous noise. The collective coordinate approach allowed for the quantitative description of non-trivial diffusive behaviour of the mean phase of the globally synchronized cluster which is an effect entirely caused by the finitude of a network and which disappears in the thermodynamic limit of infinitely many oscillators. Whereas in the deterministic case the collective coordinate approach was able to capture both the collective mean behaviour as well as the temporal behaviour of individual oscillators \cite{Gottwald15}, in the stochastic case the results are only of a statistical nature. 

 
%



\begin{thebibliography}{44}%
\makeatletter
\providecommand \@ifxundefined [1]{%
 \@ifx{#1\undefined}
}%
\providecommand \@ifnum [1]{%
 \ifnum #1\expandafter \@firstoftwo
 \else \expandafter \@secondoftwo
 \fi
}%
\providecommand \@ifx [1]{%
 \ifx #1\expandafter \@firstoftwo
 \else \expandafter \@secondoftwo
 \fi
}%
\providecommand \natexlab [1]{#1}%
\providecommand \enquote  [1]{``#1''}%
\providecommand \bibnamefont  [1]{#1}%
\providecommand \bibfnamefont [1]{#1}%
\providecommand \citenamefont [1]{#1}%
\providecommand \href@noop [0]{\@secondoftwo}%
\providecommand \href [0]{\begingroup \@sanitize@url \@href}%
\providecommand \@href[1]{\@@startlink{#1}\@@href}%
\providecommand \@@href[1]{\endgroup#1\@@endlink}%
\providecommand \@sanitize@url [0]{\catcode `\\12\catcode `\$12\catcode
  `\&12\catcode `\#12\catcode `\^12\catcode `\_12\catcode `\%12\relax}%
\providecommand \@@startlink[1]{}%
\providecommand \@@endlink[0]{}%
\providecommand \url  [0]{\begingroup\@sanitize@url \@url }%
\providecommand \@url [1]{\endgroup\@href {#1}{\urlprefix }}%
\providecommand \urlprefix  [0]{URL }%
\providecommand \Eprint [0]{\href }%
\providecommand \doibase [0]{http://dx.doi.org/}%
\providecommand \selectlanguage [0]{\@gobble}%
\providecommand \bibinfo  [0]{\@secondoftwo}%
\providecommand \bibfield  [0]{\@secondoftwo}%
\providecommand \translation [1]{[#1]}%
\providecommand \BibitemOpen [0]{}%
\providecommand \bibitemStop [0]{}%
\providecommand \bibitemNoStop [0]{.\EOS\space}%
\providecommand \EOS [0]{\spacefactor3000\relax}%
\providecommand \BibitemShut  [1]{\csname bibitem#1\endcsname}%
\let\auto@bib@innerbib\@empty
\bibitem [{\citenamefont {Pantaleone}(1998)}]{Pantaleone98}%
  \BibitemOpen
  \bibfield  {author} {\bibinfo {author} {\bibfnamefont {J.}~\bibnamefont
  {Pantaleone}},\ }\href@noop {} {\bibfield  {journal} {\bibinfo  {journal}
  {Phys. Rev. D}\ }\textbf {\bibinfo {volume} {58}},\ \bibinfo {pages} {073002}
  (\bibinfo {year} {1998})}\BibitemShut {NoStop}%
\bibitem [{\citenamefont {Sheeba}\ \emph {et~al.}(2008)\citenamefont {Sheeba},
  \citenamefont {Stefanovska},\ and\ \citenamefont
  {McClintock}}]{SheebaEtAl08}%
  \BibitemOpen
  \bibfield  {author} {\bibinfo {author} {\bibfnamefont {J.~H.}\ \bibnamefont
  {Sheeba}}, \bibinfo {author} {\bibfnamefont {A.}~\bibnamefont {Stefanovska}},
  \ and\ \bibinfo {author} {\bibfnamefont {P.~V.~E.}\ \bibnamefont
  {McClintock}},\ }\href@noop {} {\bibfield  {journal} {\bibinfo  {journal}
  {Biophysical Journal}\ }\textbf {\bibinfo {volume} {95}},\ \bibinfo {pages}
  {2722} (\bibinfo {year} {2008})}\BibitemShut {NoStop}%
\bibitem [{\citenamefont {Bhowmik}\ and\ \citenamefont
  {Shanahan}(2012)}]{BhowmikShanahan12}%
  \BibitemOpen
  \bibfield  {author} {\bibinfo {author} {\bibfnamefont {D.}~\bibnamefont
  {Bhowmik}}\ and\ \bibinfo {author} {\bibfnamefont {M.}~\bibnamefont
  {Shanahan}},\ }in\ \href@noop {} {\emph {\bibinfo {booktitle} {The 2012
  International Joint Conference on Neural Networks (IJCNN)}}}\ (\bibinfo
  {year} {2012})\ pp.\ \bibinfo {pages} {1--8}\BibitemShut {NoStop}%
\bibitem [{\citenamefont {Neda}\ \emph {et~al.}(2000)\citenamefont {Neda},
  \citenamefont {Ravasz}, \citenamefont {Brechet}, \citenamefont {Vicsek},\
  and\ \citenamefont {Barabasi}}]{NedaEtAl00}%
  \BibitemOpen
  \bibfield  {author} {\bibinfo {author} {\bibfnamefont {Z.}~\bibnamefont
  {Neda}}, \bibinfo {author} {\bibfnamefont {E.}~\bibnamefont {Ravasz}},
  \bibinfo {author} {\bibfnamefont {Y.}~\bibnamefont {Brechet}}, \bibinfo
  {author} {\bibfnamefont {T.}~\bibnamefont {Vicsek}}, \ and\ \bibinfo {author}
  {\bibfnamefont {A.~L.}\ \bibnamefont {Barabasi}},\ }\href@noop {} {\bibfield
  {journal} {\bibinfo  {journal} {Nature}\ }\textbf {\bibinfo {volume} {403}},\
  \bibinfo {pages} {849} (\bibinfo {year} {2000})}\BibitemShut {NoStop}%
\bibitem [{\citenamefont {Filatrella}\ \emph {et~al.}(2008)\citenamefont
  {Filatrella}, \citenamefont {Nielsen},\ and\ \citenamefont
  {Pedersen}}]{FilatrellaEtAl08}%
  \BibitemOpen
  \bibfield  {author} {\bibinfo {author} {\bibfnamefont {G.}~\bibnamefont
  {Filatrella}}, \bibinfo {author} {\bibfnamefont {A.~H.}\ \bibnamefont
  {Nielsen}}, \ and\ \bibinfo {author} {\bibfnamefont {N.~F.}\ \bibnamefont
  {Pedersen}},\ }\href@noop {} {\bibfield  {journal} {\bibinfo  {journal} {The
  European Physical Journal B}\ }\textbf {\bibinfo {volume} {61}},\ \bibinfo
  {pages} {485} (\bibinfo {year} {2008})}\BibitemShut {NoStop}%
\bibitem [{\citenamefont {Kuramoto}(1984)}]{Kuramoto}%
  \BibitemOpen
  \bibfield  {author} {\bibinfo {author} {\bibfnamefont {Y.}~\bibnamefont
  {Kuramoto}},\ }\href {\doibase 10.1007/978-3-642-69689-3} {\emph {\bibinfo
  {title} {Chemical {O}scillations, {W}aves, and {T}urbulence}}},\ \bibinfo
  {series} {Springer Series in Synergetics}, Vol.~\bibinfo {volume} {19}\
  (\bibinfo  {publisher} {Springer-Verlag},\ \bibinfo {address} {Berlin},\
  \bibinfo {year} {1984})\ pp.\ \bibinfo {pages} {viii+156}\BibitemShut
  {NoStop}%
\bibitem [{\citenamefont {Osipov}\ \emph {et~al.}(2007)\citenamefont {Osipov},
  \citenamefont {Kurths},\ and\ \citenamefont {Zhou}}]{OsipovEtAl}%
  \BibitemOpen
  \bibfield  {author} {\bibinfo {author} {\bibfnamefont {G.~V.}\ \bibnamefont
  {Osipov}}, \bibinfo {author} {\bibfnamefont {J.}~\bibnamefont {Kurths}}, \
  and\ \bibinfo {author} {\bibfnamefont {C.}~\bibnamefont {Zhou}},\ }\href
  {\doibase 10.1007/978-3-540-71269-5} {\emph {\bibinfo {title}
  {Synchronization in {O}scillatory {N}etworks}}},\ Springer Series in
  Synergetics\ (\bibinfo  {publisher} {Springer},\ \bibinfo {address}
  {Berlin},\ \bibinfo {year} {2007})\ p.\ \bibinfo {pages} {37c}\BibitemShut
  {NoStop}%
\bibitem [{\citenamefont {Pikovsky}\ \emph {et~al.}(2001)\citenamefont
  {Pikovsky}, \citenamefont {Rosenblum},\ and\ \citenamefont
  {Kurths}}]{PikovskyEtAl}%
  \BibitemOpen
  \bibfield  {author} {\bibinfo {author} {\bibfnamefont {A.}~\bibnamefont
  {Pikovsky}}, \bibinfo {author} {\bibfnamefont {M.}~\bibnamefont {Rosenblum}},
  \ and\ \bibinfo {author} {\bibfnamefont {J.}~\bibnamefont {Kurths}},\
  }\href@noop {} {\emph {\bibinfo {title} {{Synchronization: {A} {U}niversal
  {C}oncept in {N}onlinear {S}ciences}}}}\ (\bibinfo  {publisher} {Cambridge
  University Press},\ \bibinfo {address} {Cambridge},\ \bibinfo {year}
  {2001})\BibitemShut {NoStop}%
\bibitem [{\citenamefont {Strogatz}(2000)}]{Strogatz00}%
  \BibitemOpen
  \bibfield  {author} {\bibinfo {author} {\bibfnamefont {S.~H.}\ \bibnamefont
  {Strogatz}},\ }\href {\doibase 10.1016/S0167-2789(00)00094-4} {\bibfield
  {journal} {\bibinfo  {journal} {Physica D}\ }\textbf {\bibinfo {volume}
  {143}},\ \bibinfo {pages} {1} (\bibinfo {year} {2000})}\BibitemShut {NoStop}%
\bibitem [{\citenamefont {Acebr\'on}\ \emph {et~al.}(2005)\citenamefont
  {Acebr\'on}, \citenamefont {Bonilla}, \citenamefont {P\'erez~Vicente},
  \citenamefont {Ritort},\ and\ \citenamefont {Spigler}}]{AcebronEtAl05}%
  \BibitemOpen
  \bibfield  {author} {\bibinfo {author} {\bibfnamefont {J.~A.}\ \bibnamefont
  {Acebr\'on}}, \bibinfo {author} {\bibfnamefont {L.~L.}\ \bibnamefont
  {Bonilla}}, \bibinfo {author} {\bibfnamefont {C.~J.}\ \bibnamefont
  {P\'erez~Vicente}}, \bibinfo {author} {\bibfnamefont {F.}~\bibnamefont
  {Ritort}}, \ and\ \bibinfo {author} {\bibfnamefont {R.}~\bibnamefont
  {Spigler}},\ }\href {\doibase 10.1103/RevModPhys.77.137} {\bibfield
  {journal} {\bibinfo  {journal} {Rev. Mod. Phys.}\ }\textbf {\bibinfo {volume}
  {77}},\ \bibinfo {pages} {137} (\bibinfo {year} {2005})}\BibitemShut
  {NoStop}%
\bibitem [{\citenamefont {Arenas}\ \emph {et~al.}(2008)\citenamefont {Arenas},
  \citenamefont {Diaz-Guilera}, \citenamefont {Kurths}, \citenamefont
  {Moreno},\ and\ \citenamefont {Zhou}}]{ArenasEtAl08}%
  \BibitemOpen
  \bibfield  {author} {\bibinfo {author} {\bibfnamefont {A.}~\bibnamefont
  {Arenas}}, \bibinfo {author} {\bibfnamefont {A.}~\bibnamefont
  {Diaz-Guilera}}, \bibinfo {author} {\bibfnamefont {J.}~\bibnamefont
  {Kurths}}, \bibinfo {author} {\bibfnamefont {Y.}~\bibnamefont {Moreno}}, \
  and\ \bibinfo {author} {\bibfnamefont {C.}~\bibnamefont {Zhou}},\ }\href
  {\doibase 10.1016/j.physrep.2008.09.002} {\bibfield  {journal} {\bibinfo
  {journal} {Phys. Rep.}\ }\textbf {\bibinfo {volume} {469}},\ \bibinfo {pages}
  {93} (\bibinfo {year} {2008})}\BibitemShut {NoStop}%
\bibitem [{\citenamefont {D{\"o}rfler}\ and\ \citenamefont
  {Bullo}(2014)}]{DoerflerBullo14}%
  \BibitemOpen
  \bibfield  {author} {\bibinfo {author} {\bibfnamefont {F.}~\bibnamefont
  {D{\"o}rfler}}\ and\ \bibinfo {author} {\bibfnamefont {F.}~\bibnamefont
  {Bullo}},\ }\href@noop {} {\bibfield  {journal} {\bibinfo  {journal}
  {Automatica}\ }\textbf {\bibinfo {volume} {50}},\ \bibinfo {pages} {1539 }
  (\bibinfo {year} {2014})}\BibitemShut {NoStop}%
\bibitem [{\citenamefont {Rodrigues}\ \emph {et~al.}(2016)\citenamefont
  {Rodrigues}, \citenamefont {Peron}, \citenamefont {Ji},\ and\ \citenamefont
  {Kurths}}]{RodriguesEtAl16}%
  \BibitemOpen
  \bibfield  {author} {\bibinfo {author} {\bibfnamefont {F.~A.}\ \bibnamefont
  {Rodrigues}}, \bibinfo {author} {\bibfnamefont {T.~K.~D.}\ \bibnamefont
  {Peron}}, \bibinfo {author} {\bibfnamefont {P.}~\bibnamefont {Ji}}, \ and\
  \bibinfo {author} {\bibfnamefont {J.}~\bibnamefont {Kurths}},\ }\href@noop {}
  {\bibfield  {journal} {\bibinfo  {journal} {Physics Reports}\ }\textbf
  {\bibinfo {volume} {610}},\ \bibinfo {pages} {1 } (\bibinfo {year}
  {2016})}\BibitemShut {NoStop}%
\bibitem [{\citenamefont {Breakspear}\ \emph {et~al.}(2010)\citenamefont
  {Breakspear}, \citenamefont {Heitmann},\ and\ \citenamefont
  {Daffertshofer}}]{BreakspearEtAl10}%
  \BibitemOpen
  \bibfield  {author} {\bibinfo {author} {\bibfnamefont {M.}~\bibnamefont
  {Breakspear}}, \bibinfo {author} {\bibfnamefont {S.}~\bibnamefont
  {Heitmann}}, \ and\ \bibinfo {author} {\bibfnamefont {A.}~\bibnamefont
  {Daffertshofer}},\ }\href@noop {} {\bibfield  {journal} {\bibinfo  {journal}
  {Frontiers in Human Neuroscience}\ }\textbf {\bibinfo {volume} {4}},\
  \bibinfo {pages} {190} (\bibinfo {year} {2010})}\BibitemShut {NoStop}%
\bibitem [{\citenamefont {Cabral}\ \emph {et~al.}(2011)\citenamefont {Cabral},
  \citenamefont {Hugues}, \citenamefont {Sporns},\ and\ \citenamefont
  {Deco}}]{CabralEtAl11}%
  \BibitemOpen
  \bibfield  {author} {\bibinfo {author} {\bibfnamefont {J.}~\bibnamefont
  {Cabral}}, \bibinfo {author} {\bibfnamefont {E.}~\bibnamefont {Hugues}},
  \bibinfo {author} {\bibfnamefont {O.}~\bibnamefont {Sporns}}, \ and\ \bibinfo
  {author} {\bibfnamefont {G.}~\bibnamefont {Deco}},\ }\href@noop {} {\bibfield
   {journal} {\bibinfo  {journal} {NeuroImage}\ }\textbf {\bibinfo {volume}
  {57}},\ \bibinfo {pages} {130 } (\bibinfo {year} {2011})}\BibitemShut
  {NoStop}%
\bibitem [{\citenamefont {Faisal}\ \emph {et~al.}(2008)\citenamefont {Faisal},
  \citenamefont {Selen},\ and\ \citenamefont {Wolpert}}]{FaisalEtAl08}%
  \BibitemOpen
  \bibfield  {author} {\bibinfo {author} {\bibfnamefont {A.~A.}\ \bibnamefont
  {Faisal}}, \bibinfo {author} {\bibfnamefont {L.~P.~J.}\ \bibnamefont
  {Selen}}, \ and\ \bibinfo {author} {\bibfnamefont {D.~M.}\ \bibnamefont
  {Wolpert}},\ }\href@noop {} {\bibfield  {journal} {\bibinfo  {journal} {Nat
  Rev Neurosci}\ }\textbf {\bibinfo {volume} {9}},\ \bibinfo {pages} {292}
  (\bibinfo {year} {2008})}\BibitemShut {NoStop}%
\bibitem [{\citenamefont {Goldwyn}\ and\ \citenamefont
  {Shea-Brown}(2011)}]{GoldwynSheaBrown11}%
  \BibitemOpen
  \bibfield  {author} {\bibinfo {author} {\bibfnamefont {J.~H.}\ \bibnamefont
  {Goldwyn}}\ and\ \bibinfo {author} {\bibfnamefont {E.}~\bibnamefont
  {Shea-Brown}},\ }\href@noop {} {\bibfield  {journal} {\bibinfo  {journal}
  {PLoS Comput. Biol.}\ }\textbf {\bibinfo {volume} {7}},\ \bibinfo {pages}
  {e1002247, 9} (\bibinfo {year} {2011})}\BibitemShut {NoStop}%
\bibitem [{\citenamefont {Sakaguchi}(1988)}]{Sakaguchi88}%
  \BibitemOpen
  \bibfield  {author} {\bibinfo {author} {\bibfnamefont {H.}~\bibnamefont
  {Sakaguchi}},\ }\href@noop {} {\bibfield  {journal} {\bibinfo  {journal}
  {Progr. Theoret. Phys.}\ }\textbf {\bibinfo {volume} {79}},\ \bibinfo {pages}
  {39} (\bibinfo {year} {1988})}\BibitemShut {NoStop}%
\bibitem [{\citenamefont {Sonnenschein}\ and\ \citenamefont
  {Schimansky-Geier}(2012)}]{SonnenscheinSchimanskyGeier12}%
  \BibitemOpen
  \bibfield  {author} {\bibinfo {author} {\bibfnamefont {B.}~\bibnamefont
  {Sonnenschein}}\ and\ \bibinfo {author} {\bibfnamefont {L.}~\bibnamefont
  {Schimansky-Geier}},\ }\href@noop {} {\bibfield  {journal} {\bibinfo
  {journal} {Phys. Rev. E}\ }\textbf {\bibinfo {volume} {85}},\ \bibinfo
  {pages} {051116} (\bibinfo {year} {2012})}\BibitemShut {NoStop}%
\bibitem [{\citenamefont {Sonnenschein}\ and\ \citenamefont
  {Schimansky-Geier}(2013)}]{SonnenscheinSchimanskyGeier13}%
  \BibitemOpen
  \bibfield  {author} {\bibinfo {author} {\bibfnamefont {B.}~\bibnamefont
  {Sonnenschein}}\ and\ \bibinfo {author} {\bibfnamefont {L.}~\bibnamefont
  {Schimansky-Geier}},\ }\href@noop {} {\bibfield  {journal} {\bibinfo
  {journal} {Phys. Rev. E}\ }\textbf {\bibinfo {volume} {88}},\ \bibinfo
  {pages} {052111} (\bibinfo {year} {2013})}\BibitemShut {NoStop}%
\bibitem [{\citenamefont {Bertini}\ \emph {et~al.}(2010)\citenamefont
  {Bertini}, \citenamefont {Giacomin},\ and\ \citenamefont
  {Pakdaman}}]{BertiniEtAl10}%
  \BibitemOpen
  \bibfield  {author} {\bibinfo {author} {\bibfnamefont {L.}~\bibnamefont
  {Bertini}}, \bibinfo {author} {\bibfnamefont {G.}~\bibnamefont {Giacomin}}, \
  and\ \bibinfo {author} {\bibfnamefont {K.}~\bibnamefont {Pakdaman}},\
  }\href@noop {} {\bibfield  {journal} {\bibinfo  {journal} {Journal of
  Statistical Physics}\ }\textbf {\bibinfo {volume} {138}},\ \bibinfo {pages}
  {270} (\bibinfo {year} {2010})}\BibitemShut {NoStop}%
\bibitem [{\citenamefont {Lu\c{c}on}\ and\ \citenamefont
  {Poquet}(2015)}]{LuconPoquet15}%
  \BibitemOpen
  \bibfield  {author} {\bibinfo {author} {\bibfnamefont {E.}~\bibnamefont
  {Lu\c{c}on}}\ and\ \bibinfo {author} {\bibfnamefont {C.}~\bibnamefont
  {Poquet}},\ }\href@noop {} {\bibfield  {journal} {\bibinfo  {journal}
  {arXiv:1505.00497 [math.PR]}\ } (\bibinfo {year} {2015})}\BibitemShut
  {NoStop}%
\bibitem [{\citenamefont {Giacomin}\ and\ \citenamefont
  {Poquet}(2015)}]{GiacominPoquet15}%
  \BibitemOpen
  \bibfield  {author} {\bibinfo {author} {\bibfnamefont {G.}~\bibnamefont
  {Giacomin}}\ and\ \bibinfo {author} {\bibfnamefont {C.}~\bibnamefont
  {Poquet}},\ }\href@noop {} {\bibfield  {journal} {\bibinfo  {journal} {Braz.
  J. Probab. Stat.}\ }\textbf {\bibinfo {volume} {29}},\ \bibinfo {pages} {460}
  (\bibinfo {year} {2015})}\BibitemShut {NoStop}%
\bibitem [{\citenamefont {Lu\c{c}on}(2015)}]{Lucon15}%
  \BibitemOpen
  \bibfield  {author} {\bibinfo {author} {\bibfnamefont {E.}~\bibnamefont
  {Lu\c{c}on}},\ }in\ \href@noop {} {\emph {\bibinfo {booktitle} {From particle
  systems to partial differential equations. {II}}}},\ \bibinfo {series}
  {Springer Proc. Math. Stat.}, Vol.\ \bibinfo {volume} {129}\ (\bibinfo
  {publisher} {Springer, Cham},\ \bibinfo {year} {2015})\ pp.\ \bibinfo {pages}
  {231--251}\BibitemShut {NoStop}%
\bibitem [{\citenamefont {Lu\c{c}on}\ and\ \citenamefont
  {Stannat}(2016)}]{LuconStannat16}%
  \BibitemOpen
  \bibfield  {author} {\bibinfo {author} {\bibfnamefont {E.}~\bibnamefont
  {Lu\c{c}on}}\ and\ \bibinfo {author} {\bibfnamefont {W.}~\bibnamefont
  {Stannat}},\ }\href@noop {} {\bibfield  {journal} {\bibinfo  {journal} {Ann.
  Appl. Probab.}\ }\textbf {\bibinfo {volume} {26}},\ \bibinfo {pages} {3840}
  (\bibinfo {year} {2016})}\BibitemShut {NoStop}%
\bibitem [{\citenamefont {Dorogovtsev}\ \emph {et~al.}(2008)\citenamefont
  {Dorogovtsev}, \citenamefont {Goltsev},\ and\ \citenamefont
  {Mendes}}]{DorogovtsevEtAl08}%
  \BibitemOpen
  \bibfield  {author} {\bibinfo {author} {\bibfnamefont {S.~N.}\ \bibnamefont
  {Dorogovtsev}}, \bibinfo {author} {\bibfnamefont {A.~V.}\ \bibnamefont
  {Goltsev}}, \ and\ \bibinfo {author} {\bibfnamefont {J.~F.~F.}\ \bibnamefont
  {Mendes}},\ }\href@noop {} {\bibfield  {journal} {\bibinfo  {journal} {Rev.
  Mod. Phys.}\ }\textbf {\bibinfo {volume} {80}},\ \bibinfo {pages} {1275}
  (\bibinfo {year} {2008})}\BibitemShut {NoStop}%
\bibitem [{\citenamefont {Ott}\ and\ \citenamefont
  {Antonsen}(2008)}]{OttAntonson08}%
  \BibitemOpen
  \bibfield  {author} {\bibinfo {author} {\bibfnamefont {E.}~\bibnamefont
  {Ott}}\ and\ \bibinfo {author} {\bibfnamefont {T.~M.}\ \bibnamefont
  {Antonsen}},\ }\href {\doibase 10.1063/1.2930766} {\bibfield  {journal}
  {\bibinfo  {journal} {Chaos}\ }\textbf {\bibinfo {volume} {18}},\ \bibinfo
  {pages} {037113, 6} (\bibinfo {year} {2008})}\BibitemShut {NoStop}%
\bibitem [{\citenamefont {Pikovsky}\ and\ \citenamefont
  {Rosenblum}(2015)}]{RosenblumPikovsky15}%
  \BibitemOpen
  \bibfield  {author} {\bibinfo {author} {\bibfnamefont {A.}~\bibnamefont
  {Pikovsky}}\ and\ \bibinfo {author} {\bibfnamefont {M.}~\bibnamefont
  {Rosenblum}},\ }\href@noop {} {\bibfield  {journal} {\bibinfo  {journal}
  {Chaos}\ }\textbf {\bibinfo {volume} {25}},\ \bibinfo {pages} {097616}
  (\bibinfo {year} {2015})}\BibitemShut {NoStop}%
\bibitem [{\citenamefont {Hildebrand}\ \emph {et~al.}(2007)\citenamefont
  {Hildebrand}, \citenamefont {Buice},\ and\ \citenamefont
  {Chow}}]{HildebrandtEtAl07}%
  \BibitemOpen
  \bibfield  {author} {\bibinfo {author} {\bibfnamefont {E.~J.}\ \bibnamefont
  {Hildebrand}}, \bibinfo {author} {\bibfnamefont {M.~A.}\ \bibnamefont
  {Buice}}, \ and\ \bibinfo {author} {\bibfnamefont {C.~C.}\ \bibnamefont
  {Chow}},\ }\href {\doibase 10.1103/PhysRevLett.98.054101} {\bibfield
  {journal} {\bibinfo  {journal} {Phys. Rev. Lett.}\ }\textbf {\bibinfo
  {volume} {98}},\ \bibinfo {pages} {054101} (\bibinfo {year}
  {2007})}\BibitemShut {NoStop}%
\bibitem [{\citenamefont {Buice}\ and\ \citenamefont
  {Chow}(2007)}]{BuiceChow07}%
  \BibitemOpen
  \bibfield  {author} {\bibinfo {author} {\bibfnamefont {M.~A.}\ \bibnamefont
  {Buice}}\ and\ \bibinfo {author} {\bibfnamefont {C.~C.}\ \bibnamefont
  {Chow}},\ }\href {\doibase 10.1103/PhysRevE.76.031118} {\bibfield  {journal}
  {\bibinfo  {journal} {Phys. Rev. E}\ }\textbf {\bibinfo {volume} {76}},\
  \bibinfo {pages} {031118} (\bibinfo {year} {2007})}\BibitemShut {NoStop}%
\bibitem [{\citenamefont {Hong}\ \emph {et~al.}(2007)\citenamefont {Hong},
  \citenamefont {Ha},\ and\ \citenamefont {Park}}]{HongEtAl07}%
  \BibitemOpen
  \bibfield  {author} {\bibinfo {author} {\bibfnamefont {H.}~\bibnamefont
  {Hong}}, \bibinfo {author} {\bibfnamefont {M.}~\bibnamefont {Ha}}, \ and\
  \bibinfo {author} {\bibfnamefont {H.}~\bibnamefont {Park}},\ }\href {\doibase
  10.1103/PhysRevLett.98.258701} {\bibfield  {journal} {\bibinfo  {journal}
  {Phys. Rev. Lett.}\ }\textbf {\bibinfo {volume} {98}},\ \bibinfo {pages}
  {258701} (\bibinfo {year} {2007})}\BibitemShut {NoStop}%
\bibitem [{\citenamefont {Tang}(2011)}]{Tang11}%
  \BibitemOpen
  \bibfield  {author} {\bibinfo {author} {\bibfnamefont {L.-H.}\ \bibnamefont
  {Tang}},\ }\href {http://stacks.iop.org/1742-5468/2011/i=01/a=P01034}
  {\bibfield  {journal} {\bibinfo  {journal} {Journal of Statistical Mechanics:
  Theory and Experiment}\ }\textbf {\bibinfo {volume} {2011}},\ \bibinfo
  {pages} {P01034} (\bibinfo {year} {2011})}\BibitemShut {NoStop}%
\bibitem [{\citenamefont {Gottwald}(2015)}]{Gottwald15}%
  \BibitemOpen
  \bibfield  {author} {\bibinfo {author} {\bibfnamefont {G.~A.}\ \bibnamefont
  {Gottwald}},\ }\href@noop {} {\bibfield  {journal} {\bibinfo  {journal}
  {Chaos}\ }\textbf {\bibinfo {volume} {25}},\ \bibinfo {pages} {053111, 12}
  (\bibinfo {year} {2015})}\BibitemShut {NoStop}%
\bibitem [{\citenamefont {Watanabe}\ and\ \citenamefont
  {Strogatz}(1993)}]{WatanabeStrogatz93}%
  \BibitemOpen
  \bibfield  {author} {\bibinfo {author} {\bibfnamefont {S.}~\bibnamefont
  {Watanabe}}\ and\ \bibinfo {author} {\bibfnamefont {S.~H.}\ \bibnamefont
  {Strogatz}},\ }\href@noop {} {\bibfield  {journal} {\bibinfo  {journal}
  {Phys. Rev. Lett.}\ }\textbf {\bibinfo {volume} {70}},\ \bibinfo {pages}
  {2391} (\bibinfo {year} {1993})}\BibitemShut {NoStop}%
\bibitem [{\citenamefont {Marvel}\ \emph {et~al.}(2009)\citenamefont {Marvel},
  \citenamefont {Mirollo},\ and\ \citenamefont {Strogatz}}]{MarvelEtAl09}%
  \BibitemOpen
  \bibfield  {author} {\bibinfo {author} {\bibfnamefont {S.~A.}\ \bibnamefont
  {Marvel}}, \bibinfo {author} {\bibfnamefont {R.~E.}\ \bibnamefont {Mirollo}},
  \ and\ \bibinfo {author} {\bibfnamefont {S.~H.}\ \bibnamefont {Strogatz}},\
  }\href {\doibase 10.1063/1.3247089} {\bibfield  {journal} {\bibinfo
  {journal} {Chaos}\ }\textbf {\bibinfo {volume} {19}},\ \bibinfo {pages}
  {043104, 11} (\bibinfo {year} {2009})}\BibitemShut {NoStop}%
\bibitem [{\citenamefont {Pikovsky}\ and\ \citenamefont
  {Rosenblum}(2011)}]{PikovskyRosenblum11}%
  \BibitemOpen
  \bibfield  {author} {\bibinfo {author} {\bibfnamefont {A.}~\bibnamefont
  {Pikovsky}}\ and\ \bibinfo {author} {\bibfnamefont {M.}~\bibnamefont
  {Rosenblum}},\ }\href@noop {} {\bibfield  {journal} {\bibinfo  {journal}
  {Physica D}\ }\textbf {\bibinfo {volume} {240}},\ \bibinfo {pages} {872 }
  (\bibinfo {year} {2011})}\BibitemShut {NoStop}%
\bibitem [{\citenamefont {Pinto}\ and\ \citenamefont {Saa}(2015)}]{PintoSaa15}%
  \BibitemOpen
  \bibfield  {author} {\bibinfo {author} {\bibfnamefont {R.~S.}\ \bibnamefont
  {Pinto}}\ and\ \bibinfo {author} {\bibfnamefont {A.}~\bibnamefont {Saa}},\
  }\href@noop {} {\bibfield  {journal} {\bibinfo  {journal} {Phys. Rev. E (3)}\
  }\textbf {\bibinfo {volume} {92}},\ \bibinfo {pages} {062801, 6} (\bibinfo
  {year} {2015})}\BibitemShut {NoStop}%
\bibitem [{\citenamefont {Brede}\ and\ \citenamefont
  {Kalloniatis}(2016)}]{BredeKalloniatis16}%
  \BibitemOpen
  \bibfield  {author} {\bibinfo {author} {\bibfnamefont {M.}~\bibnamefont
  {Brede}}\ and\ \bibinfo {author} {\bibfnamefont {A.~C.}\ \bibnamefont
  {Kalloniatis}},\ }\href@noop {} {\bibfield  {journal} {\bibinfo  {journal}
  {Phys. Rev. E}\ }\textbf {\bibinfo {volume} {93}},\ \bibinfo {pages} {062315}
  (\bibinfo {year} {2016})}\BibitemShut {NoStop}%
\bibitem [{\citenamefont {Scott}(2003)}]{Scott}%
  \BibitemOpen
  \bibfield  {author} {\bibinfo {author} {\bibfnamefont {A.}~\bibnamefont
  {Scott}},\ }\href@noop {} {\emph {\bibinfo {title} {Nonlinear {S}cience}}},\
  \bibinfo {edition} {2nd}\ ed.,\ \bibinfo {series} {Oxford Texts in Applied
  and Engineering Mathematics}, Vol.~\bibinfo {volume} {8}\ (\bibinfo
  {publisher} {Oxford University Press},\ \bibinfo {address} {Oxford},\
  \bibinfo {year} {2003})\ pp.\ \bibinfo {pages} {xxiv+480},\ \bibinfo {note}
  {emergence and {D}ynamics of {C}oherent {S}tructures}\BibitemShut {NoStop}%
\bibitem [{\citenamefont {Gottwald}\ and\ \citenamefont
  {Kramer}(2004)}]{GottwaldKramer04}%
  \BibitemOpen
  \bibfield  {author} {\bibinfo {author} {\bibfnamefont {G.~A.}\ \bibnamefont
  {Gottwald}}\ and\ \bibinfo {author} {\bibfnamefont {L.}~\bibnamefont
  {Kramer}},\ }\href {\doibase http://dx.doi.org/10.1063/1.1772552} {\bibfield
  {journal} {\bibinfo  {journal} {Chaos}\ }\textbf {\bibinfo {volume} {14}},\
  \bibinfo {pages} {855} (\bibinfo {year} {2004})}\BibitemShut {NoStop}%
\bibitem [{\citenamefont {Menon}\ and\ \citenamefont
  {Gottwald}(2005)}]{MenonGottwald05}%
  \BibitemOpen
  \bibfield  {author} {\bibinfo {author} {\bibfnamefont {S.~N.}\ \bibnamefont
  {Menon}}\ and\ \bibinfo {author} {\bibfnamefont {G.~A.}\ \bibnamefont
  {Gottwald}},\ }\href {\doibase 10.1103/PhysRevE.71.066201} {\bibfield
  {journal} {\bibinfo  {journal} {Phys. Rev. E}\ }\textbf {\bibinfo {volume}
  {71}},\ \bibinfo {pages} {066201} (\bibinfo {year} {2005})}\BibitemShut
  {NoStop}%
\bibitem [{\citenamefont {Menon}\ and\ \citenamefont
  {Gottwald}(2007)}]{MenonGottwald07}%
  \BibitemOpen
  \bibfield  {author} {\bibinfo {author} {\bibfnamefont {S.~N.}\ \bibnamefont
  {Menon}}\ and\ \bibinfo {author} {\bibfnamefont {G.~A.}\ \bibnamefont
  {Gottwald}},\ }\href@noop {} {\bibfield  {journal} {\bibinfo  {journal}
  {Phys. Rev. E}\ }\textbf {\bibinfo {volume} {75}},\ \bibinfo {pages} {016209}
  (\bibinfo {year} {2007})}\BibitemShut {NoStop}%
\bibitem [{\citenamefont {Menon}\ and\ \citenamefont
  {Gottwald}(2009)}]{MenonGottwald09}%
  \BibitemOpen
  \bibfield  {author} {\bibinfo {author} {\bibfnamefont {S.~N.}\ \bibnamefont
  {Menon}}\ and\ \bibinfo {author} {\bibfnamefont {G.~A.}\ \bibnamefont
  {Gottwald}},\ }\href@noop {} {\bibfield  {journal} {\bibinfo  {journal}
  {Physica D}\ }\textbf {\bibinfo {volume} {238}},\ \bibinfo {pages} {461}
  (\bibinfo {year} {2009})}\BibitemShut {NoStop}%
\bibitem [{\citenamefont {Cox}\ and\ \citenamefont
  {Gottwald}(2006)}]{CoxGottwald06}%
  \BibitemOpen
  \bibfield  {author} {\bibinfo {author} {\bibfnamefont {S.~M.}\ \bibnamefont
  {Cox}}\ and\ \bibinfo {author} {\bibfnamefont {G.~A.}\ \bibnamefont
  {Gottwald}},\ }\href {\doibase http://dx.doi.org/10.1016/j.physd.2006.03.007}
  {\bibfield  {journal} {\bibinfo  {journal} {Physica D}\ }\textbf {\bibinfo
  {volume} {216}},\ \bibinfo {pages} {307 } (\bibinfo {year}
  {2006})}\BibitemShut {NoStop}%
\end{thebibliography}

%

\end{document}